\documentclass[aps,prb,reprint,longbibliography,superscriptaddress]{revtex4-2}
\usepackage[utf8]{inputenc}
\usepackage{amsmath,amssymb}
\usepackage{graphicx}   
\DeclareMathOperator{\Tr}{Tr}
\usepackage{hyperref}     
\usepackage{graphicx}     

\hypersetup{
	colorlinks=true,
	linkcolor=blue,
	citecolor=blue,
	urlcolor=blue
}
\usepackage{graphicx}
\usepackage{epstopdf}
\begin{document}
\title{Anti-Kibble-Zurek behavior driven by dichotomous Markov noise in a quantum Ising chain: Generalized stochastic resonance in the defect density}
\author{Mingwei Yang}
\affiliation{College of Physics, Sichuan University, 610064, Chengdu, People’s Republic of China}
\author{Peng Li}
\email{lipeng@scu.edu.cn}   
\affiliation{College of Physics, Sichuan University, 610064, Chengdu, People’s Republic of China}
\begin{abstract}
The anti-Kibble-Zurek behavior may arise under noise, adding a noise-induced contribution to the defect density beyond the conventional Kibble-Zurek term. Most investigations of such behavior have so far been restricted to Gaussian noise. For Gaussian colored noise (GCN), the power spectrum dictates a monotonic variation of this term with the correlation time. Here, we focus on dichotomous Markov noise (DMN), a prototypical non-Gaussian colored noise, in the quantum Ising chain. We show that the noise-induced defect density, taken as the response function, exhibits generalized stochastic resonance: it is nonmonotonic in the correlation time, growing in the fast-noise regime and decreasing in the slow-noise regime. This resonance stems from the anomalous AKZ behavior in the slow-noise regime, where the DMN-induced trend is nonmonotonic, in contrast to the monotonic GCN case. The underlying mechanism lies in the sparse realizations of DMN in the slow-noise regime, a feature not captured by the power spectrum. Our work thereby reveals a fundamental distinction between Gaussian and non-Gaussian noises in the nonequilibrium critical quench dynamics of the quantum Ising chain.
\end{abstract}
\date{\today}
\maketitle
\section{Introduction}
	The Kibble-Zurek mechanism (KZM) provides a fundamental paradigm for the critical dynamics of phase transitions from cosmology to condensed matter systems~\cite{Kibble1976,Zurek1985,Dziarmaga2010}. According to KZM, gap closing at the critical point drives non-adiabatic evolution during a slow quench in the thermodynamic limit. The KZM has been verified in classical continuous phase transitions across a wide range of systems~\cite{Ulm2013,Pyka2013}, and it also governs quantum phase transitions, with its quantum counterpart (QKZM) having been investigated both theoretically~\cite{Zurek2005,Dziarmaga2005} and experimentally~\cite{Cui2016,Keesling2019,King2022,Miessen2024}. Quantitatively, the defect density $n$ scales with the quench time $\tau_q$ as $n \propto \tau_q^{-\beta}$, with $\beta = d\nu/(1+z\nu)$, where $d$ is the spatial dimension, and $\nu$ and $z$ are the correlation-length and dynamical critical exponents, respectively~\cite{Zurek2005,Polkovnikov2005,Dziarmaga2005}.
    
However, Kibble–Zurek scaling is not guaranteed~\cite{Jafari2026,Jafari2026arxiv}.	
   Notably, a trend that violates the conventional KZM scaling has been observed in ferroelectric phase transition: slower quenches produce an increasing number of topological defects near the adiabatic limit, a phenomenon now referred to as anti-Kibble-Zurek (AKZ) behavior~\cite{Griffin2012}. Such a behavior has also been observed experimentally in trapped-ion~\cite{Ai2021} and D-Wave annealer platforms~\cite{King2022}. Theoretically, it has been shown that the AKZ can be induced by adding the Gaussian white noise to the driven control parameter in the transverse-field Ising chain~\cite{Dutta2016}. The analysis was subsequently extended from weak to strong noise regimes via an analytical approach~\cite{Iwamura2024}.  

    Recently, the effect of temporal correlations in Gaussian colored noise (GCN) with an Ornstein--Uhlenbeck correlation~\cite{Oksendal2003} on the defect density was studied for a transverse-field quantum XY chain~\cite{Sadeghizade2025}, where the defect density decomposes into a Kibble--Zurek term and a noise contribution: \(n = \frac{1}{2\pi}\,\tau_q^{-1/2} + \Delta n,\) with \(\Delta n = b\,W^2 \tau_q^2 \arctan\!\left(\frac{c\,\tau_n}{\tau_q}\right),\) where $b$ and $c$ are constants determined numerically, $\tau_n$ is the correlation time, and $W^2$ is the total noise power. This expression, obtained via a power-spectrum-based analysis~\cite{Sadeghizade2025}, shows that $\Delta n$ increases monotonically with $\tau_n$. Note that these conclusions are based on Gaussian noise, regardless of whether it is white or temporally correlated.
	
	In practical solid-state quantum devices, non-Gaussian noise is ubiquitous. A paradigmatic example is dichotomous Markov noise (DMN) arising from two-level charge traps, which provides a minimal model of non-Gaussian colored noise with finite correlation time~\cite{Paladino2014,Galperin2006,Bergli2009,Sun2012,Catelani2011}. Indeed, Doob's theorem states that the unique stationary Gaussian Markov process with a finite correlation time is the Ornstein--Uhlenbeck process~\cite{Doob1942}. Consequently, DMN and GCN with the same Ornstein--Uhlenbeck correlation are not equivalent: the non-Gaussian character of DMN cannot be eliminated without losing a finite correlation time~\cite{Bena2006}. 
	This renders DMN indispensable for the study of higher-order statistics, as exemplified by stochastic resonance~\cite{Zhang2018} and noise-induced quantum non-Markovianity~\cite{Daniotti2018}.

    These developments motivate us to investigate whether the AKZ behavior driven by DMN differs from its Gaussian-colored-noise counterpart in the transverse-field Ising chain. Notably, GCN with the Ornstein--Uhlenbeck correlation and DMN share the same Lorentzian power spectrum. Would the power spectrum alone then predict identical AKZ behavior for both? Our work, however, demonstrates that this implication breaks down for DMN. Specifically, while GCN yields a monotonic increase of $\Delta n$ with $\tau_n$, DMN leads to a non-monotonic dependence: $\Delta n$ first grows and then is suppressed as the noise slows down.
    By identifying $\Delta n$ as the response function, we can view this non-monotonic dependence as a generalized stochastic resonance (GSR)~\cite{Hu1993,Fulinski1995,Collins1996,Wellens2004,Lofstedt1994,Han2012,Tian2025,Tian2026}. This resonance stems from the behavior of DMN in the slow-noise regime, which constitutes an anomalous anti-Kibble-Zurek behavior.
    
   We attribute this behavior to the {\it sparse realizations} of DMN, namely rare sign-switching events near level crossings~\cite{Luo2017}. In this regime, the DMN-induced noise term scales as $\Delta n \propto \tau_q/\tau_n$ and therefore decreases with $\tau_n$. This behavior is dictated by the detailed statistical properties of the noise, whereas the power spectrum captures only second-order statistics. Thus, for non-Gaussian colored noise, the power spectrum alone is generally insufficient to characterize the AKZ behavior; specific statistical features—such as the {\it sparse realizations}—are often crucial.

The outline of this paper is as follows. Section~\ref{sec:noise_model} presents the master equations for GCN and DMN. Section~\ref{AKZ} discusses the AKZ behavior of GCN and DMN. Finally, we summarize our findings in Sec.~\ref{sum}.   
\section{Noise-Averaged Master Equations }
\label{sec:noise_model}
\subsection{General Master Equations for GCN and DMN}
	
	We consider the general stochastic Hamiltonian
\begin{equation}
    H(t,\eta(t)) = H_0(t) + \eta(t) H_1(t), \label{SH}
\end{equation}
where $\eta(t)$ denotes the noise, $H_0(t)$ is the noise-free part, and $H_1(t)$ describes the coupling to the noise. The state $|\psi_\eta(t)\rangle$ evolves according to the stochastic Schr\"odinger equation (SSE)~\cite{Breuer2007,Biele2012,Barchielli2013,Barchielli2010,DeKeijzer2025a,DeKeijzer2025b} generated by $H(t,\eta(t))$:
\begin{equation}
    i\frac{d}{dt} |\psi_\eta(t)\rangle = H(t,\eta(t)) |\psi_\eta(t)\rangle.
    \label{eq:SSE}
\end{equation}
Differentiating the pure-state representation
\begin{equation}
    \rho_\eta(t)=|\psi_\eta(t)\rangle\langle\psi_\eta(t)|
\end{equation}
and using Eq.~\eqref{eq:SSE}, we obtain the von Neumann equation for $\rho_\eta(t)$:
\begin{equation}
    \frac{d}{dt} \rho_{\eta}(t) = -i \bigl[ H(t,\eta(t)), \rho_{\eta}(t) \bigr].
    \label{eq:vNe}
\end{equation}
Averaging Eq.~\eqref{eq:vNe} over noise realizations leads to the following master equation for the noise-averaged density operator, $\rho(t) = \langle \rho_{\eta}(t) \rangle$:
\begin{equation}
    \frac{d}{dt}\rho(t) = -i\bigl[H_0(t),\rho(t)\bigr] - i\bigl[H_1(t),\langle \eta(t)\rho_{\eta}(t)\rangle\bigr].
    \label{NADM}
\end{equation}
Here, $\langle\cdot\rangle$ denotes the average over noise realizations. This notation is used throughout the paper.

Since the concrete master equation relies on the type of noise, we address two types, GCN and DMN, in the following.
	
For GCN, we consider the Ornstein--Uhlenbeck correlation function
\begin{equation}
    R(t) = \langle \eta(0) \eta(t) \rangle = A^2 \exp\left(-\frac{|t|}{\tau_n}\right), \label{gauss_corr}
\end{equation}
where $\tau_n$ is the correlation time and $A^2$ is the noise power. The corresponding power spectrum is obtained from the Fourier transform~\cite{Wiener1930,Khintchine1934,Wang1945,Mallick2007}:
\begin{equation}
    S(\omega) = \frac{\tau_n A^2}{\pi(\omega^2\tau_n^2 + 1)}, \label{spec_gauss}
\end{equation}
which has a Lorentzian form.

Applying Novikov's theorem~\cite{Novikov1965} for GCN to Eq.~\eqref{NADM} yields
\begin{equation}
    \begin{aligned}
        \frac{d}{dt}\rho(t) &= -i[H_0(t),\rho(t)] \\
        &\quad -A^2\left[H_1(t),\int_{t_i}^t e^{-\frac{t-s}{\tau_n}}[H_1(s),\rho(s)]\,ds\right].
    \end{aligned}\label{CNME}
\end{equation}

The DMN is generated by a two-sided Poisson process \(N(t)\) of rate \(\lambda\), via
\begin{equation}
    \eta(t) = \eta(0)(-1)^{N(t)},
    \label{eq:POisson}
\end{equation}
which produces two equally probable values, e.g., \(\pm A\), as shown in Appendix~\ref{app:FCS}. The associated correlation time is
\begin{equation}
    \tau_n = \frac{1}{2\lambda},
    \label{correlation_time}
\end{equation}
and the correlation function is identical to Eq.~\eqref{gauss_corr}. Accordingly, the power spectrum coincides with Eq.~\eqref{spec_gauss} and is likewise Lorentzian.

Upon substituting the Shapiro--Loginov formula~\cite{Shapiro1978,Goychuk1995,Kosov2018} into Eq.~\eqref{NADM}, the master equation for DMN reduces to~\cite{Kiely2021}:
\begin{equation}
    \begin{aligned}
        &\frac{d}{dt}\rho(t) = -i [H_0,\rho(t)] - A^2\times \\
        &\left[ H_1(t), \int_{t_i}^t ds \, e^{-2\lambda (t-s)} \, U_0(t,s) \, [H_1(s),\rho(s)] \, U_0^\dagger(t,s) \right],\label{DNME}
    \end{aligned}
\end{equation}
with the noise-free time evolution operator $U_0(t,s)$ given by
\begin{equation}
    U_0(t,s) = \mathcal{T} \exp\left[-i \int_s^t d\tau \, H_0(\tau)\right],
\end{equation}
where \(\mathcal{T}\) denotes the Dyson time-ordering operator.
\subsection{Transverse-field quantum Ising chain}	
	
	The stochastic Hamiltonian of the transverse-field quantum Ising chain is given by
\begin{equation}
    H_{\eta}(t) = -\left[\sum_{j=1}^{N} \sigma_{j}^{x}\sigma_{j+1}^{x} + g(t)\sum_{j=1}^{N}\sigma_{j}^{z}\right].\label{ISING}
\end{equation}
The time-dependent parameter is defined as \begin{equation}g(t) = g_0(t) +\eta(t),
\end{equation} where \(g_0(t)\) is the deterministic linear ramp and \(\eta(t)\) denotes a real-valued noise realization. Here, \begin{equation}
g_0(t) = g_i + (g_f - g_i)t/
\tau_q\label{17}
\end{equation} 
varies linearly from \(g_i=-2\) at \(t_i = 0\) to \(g_f=0\) at \(t_f = \tau_q\), where \(\tau_q\) is the quench time; it thus crosses the critical point \(g_c = -1\).
The Pauli matrices are denoted by \(\sigma_{j}^{a}\) \((a = x,z)\), and the total number of lattice sites \(N\) is assumed to be even. We impose the periodic boundary condition \(\sigma_{N + j}^{a} = \sigma_{j}^{a}\).
	
By mapping the spin operators onto free fermions via the Jordan-Wigner transformation~\cite{Franchini2017}, we arrive at
\begin{equation}
		\begin{split}
			H_{\eta}(t) = &-\sum_{j}^{N} \left( c_{j}^{\dagger}c_{j+1}^{\dagger} - c_{j}c_{j+1} + c_{j}^{\dagger}c_{j+1} - c_{j}c_{j+1}^{\dagger} \right) \\
			&-  g(t) \sum_{j}^{N} \left( 2c_{j}^{\dagger}c_{j} - 1 \right).
		\end{split}\label{eq:4}
\end{equation}
In momentum space, the Fourier transform consistent with the antiperiodic boundary condition \(c_{N+1} = -c_1\) is
\begin{equation}
		c_j = \frac{e^{-i\pi/4}}{\sqrt{N}} \sum_{k} c_k e^{i k j},
		\label{eq:Fourier}
\end{equation}
where \(k = \pm(2m-1)\pi/N\) with \(m = 1, 2, \dots, N/2\).

Introducing the Nambu spinor \(\Psi_{k}^{\dagger} \equiv (c_{k}^{\dagger}\; c_{-k})\) and projecting onto the even parity subspace yields
	\begin{equation}
		H_{\eta}(t) = \sum_{k > 0} \Psi_{k}^{\dagger} h_{k,\eta}(t) \Psi_{k}\label{DT},
	\end{equation}
where
	\begin{equation}
		h_{k,\eta} (t)= 2\bigl[g(t) - \cos(k)\bigr]\hat{\sigma}^z + 2\sin(k)\hat{\sigma}^x.
	\end{equation}
Substituting Eq.~\eqref{DT} into Eq.~\eqref{eq:vNe} yields
	\begin{equation}
		\frac{d}{dt} \rho_{\eta}(t) = -i \bigl[ H_{\eta}(t), \rho_{\eta}(t) \bigr] \label{eq:FN}.
	\end{equation}

Since the system decomposes into a sum of decoupled modes labeled by $k$~\cite{Barouch1970}, we choose the initial state to be a direct product over $k$, so that the density operator retains its product form at all times:
\begin{equation}
    \rho_{\eta}(t) = \bigotimes_{k > 0} \rho_{k,\eta}(t),
\end{equation}
where \(\rho_{k,\eta}(t)\) is the reduced density operator acting on the two-dimensional Hilbert space of the $k$-th mode. 

Projecting Eq.~\eqref{eq:FN} onto each $k$ subspace and tracing out the others gives the closed evolution for each mode, as shown in Appendix~\ref{app:derivation}. In the pseudospin basis \(|0\rangle_k\) and \(c_k^\dagger c_{-k}^\dagger |0\rangle_k\), the operator equation~\eqref{app:final} reduces to the following matrix equation for the representation \(O_{k,\eta}(t)\) of \(\rho_{k,\eta}(t)\):
\begin{equation}
    \frac{d}{dt} O_{k,\eta}(t) = -i \bigl[ h_{k,\eta}(t), O_{k,\eta}(t) \bigr],\label{eq:REP}
\end{equation}
where \(h_{k,\eta}(t)\) and \(O_{k,\eta}(t)\) are \(2 \times 2\) matrices. Explicitly, we have
\begin{equation}
    h_{k,\eta}(t) = h'_{0,k}(t) + \eta(t) h''_{1,k}(t),\label{eq:HAM0}
\end{equation}
with
\begin{equation}
    h'_{0,k}(t)=2\bigl[g_0(t) - \cos(k)\bigr]\hat{\sigma}^z + 2\sin(k)\hat{\sigma}^x,\label{eq:HAM1}
\end{equation}
and $h''_{1,k}(t)=2\hat{\sigma}^z$. In the absence of noise, i.e., $\eta(t)=0$, Eq.~\eqref{eq:REP} reduces to
\begin{equation}
    \frac{d}{dt} O_{k,0}(t) = -i \bigl[ h'_{0,k}(t), O_{k,0}(t) \bigr],
    \label{eq:REP0}
\end{equation}
with $O_{k,0}(t)\equiv O_{k,\eta}(t)|_{\eta=0}$.
With the definitions
\begin{equation}
    t' = 4\tau_{q}\sin (k)\left(\frac{t}{\tau_{q}} + \frac{g_i - \cos(k)}{g_f - g_i}\right)
\end{equation}
and
\begin{equation}
    \tilde{\eta}_k(t') = \frac{\eta(t)}{\sin(k)},\label{eq:noise def}
\end{equation}
together with the  sweep velocity \(v_k = \frac{g_f - g_i}{4\tau_{q}\sin^{2}(k)}\), Eq.~\eqref{eq:HAM0} becomes
\begin{equation}
    h_{k,\tilde{\eta}}(t') = \frac{1}{2}[v_k t' + \tilde{\eta}_k(t')]\sigma_z + \frac{1}{2} \sigma_x,\label{eq:HAM}
\end{equation}
which is in fact a two-level Landau--Zener problem~\cite{Landau1932} with diagonal energy fluctuation~\cite{Kayanuma1984}. Thus, Eq.~\eqref{eq:REP} becomes
\begin{equation}
    \frac{d}{dt'} O_{k,\tilde{\eta}}(t') = -i \bigl[ h_{k,\tilde{\eta}}(t'), O_{k,\tilde{\eta}}(t') \bigr].
    \label{eq:REP2}
\end{equation}
As seen from Eqs.~\eqref{eq:HAM} and \eqref{eq:REP2}, the dynamics generated by the stochastic Hamiltonian~\eqref{ISING} reduces to a collection of decoupled noisy Landau--Zener processes. We then define the effective Landau--Zener characteristic time for each $k$-mode,
\begin{equation}
    \tau_{\mathrm{LZ}}^{(k)} = 2\tau_{q}\sin^{2}(k).
    \label{eq:LZ}
\end{equation}
	
Averaging Eq.~\eqref{eq:REP} over noise realizations leads to the following master equation for the noise-averaged density matrix of mode $k$, \(O_k(t) = \langle O_{k,\eta}(t) \rangle\):
\begin{equation}
    \frac{d}{dt} O_k(t) = -i\bigl[h'_{0,k}(t),O_k(t)\bigr]
    - i\bigl[h''_{1,k}(t),\langle \eta(t)O_{k,\eta}(t)\rangle\bigr]. \label{eq:9}
\end{equation}
	
	Eq.~\eqref{eq:9} is the concrete form of the general master Eq.~\eqref{NADM} for each $k$ mode of the Ising chain, represented in the pseudospin basis. Consequently, for GCN, Eq.~\eqref{CNME} becomes
\begin{equation}
    \begin{aligned}
        \frac{d}{dt}O_k(t) = &-i[h'_{0,k}(t),O_k(t)] \\
        &-A^2\bigl[\tilde{h}_{1,k}(t),\int_{t_i}^t e^{-\frac{t-s}{\tau_n}}[h''_{1,k}(s),O_k(s)]\,ds\bigr],
    \end{aligned}
    \label{GAUSS}
\end{equation}
while for dichotomous Markov noise Eq.~\eqref{DNME} yields
\begin{equation}
    \begin{aligned}
        &\frac{d}{dt}O_k(t) = -i[h'_{0,k}(t),O_k(t)] - A^2\times \\
        &\Bigl[\tilde{h}_{1,k}(t),\int_{t_i}^t\!ds\,e^{-2\lambda(t-s)}u_{0,k}(t,s)
        [h''_{1,k}(s),O_k(s)]u_{0,k}^\dagger(t,s)\Bigr],
    \end{aligned}
    \label{DMN}
\end{equation}
where $u_{0,k}(t,s)=\mathcal{T}\exp[-i\int_s^t d\tau\,h'_{0,k}(\tau)]$ is the matrix representation of the noise-free evolution operator for mode $k$.

By numerically solving Eqs.~\eqref{GAUSS} and~\eqref{DMN}, the density of excitations is obtained from the expectation values of the noise-averaged density matrix $O_k(t_f)$:
\begin{equation}
    n = \frac{2}{N}\sum_{k > 0}\langle \epsilon_k(t_f)|O_k(t_f)|\epsilon_k(t_f)\rangle,
\end{equation}
where \(\left|\epsilon_k(t_f)\right\rangle=
\begin{pmatrix}
\sin(k/2)\\
\cos(k/2)
\end{pmatrix}\) denotes the excited eigenstate of the Hamiltonian \( h'_{0,k}\) at the end of the quench, \(t = t_f\)~\cite{Sadeghizade2025}. The noise-induced defect density follows accordingly,
\begin{equation}
    \Delta n = \frac{2}{N}\sum_{k > 0}\langle \epsilon_k(t_f)|\bigl[O_k(t_f) - O_{k,0}(t_f)\bigr]|\epsilon_k(t_f)\rangle.
    \label{eq:Delta_n}
\end{equation}

\section{Noise-Driven Anti-Kibble-Zurek Behavior}\label{AKZ}
	\begin{figure*}[htbp]
		\centering
		\includegraphics[width=0.98\linewidth]{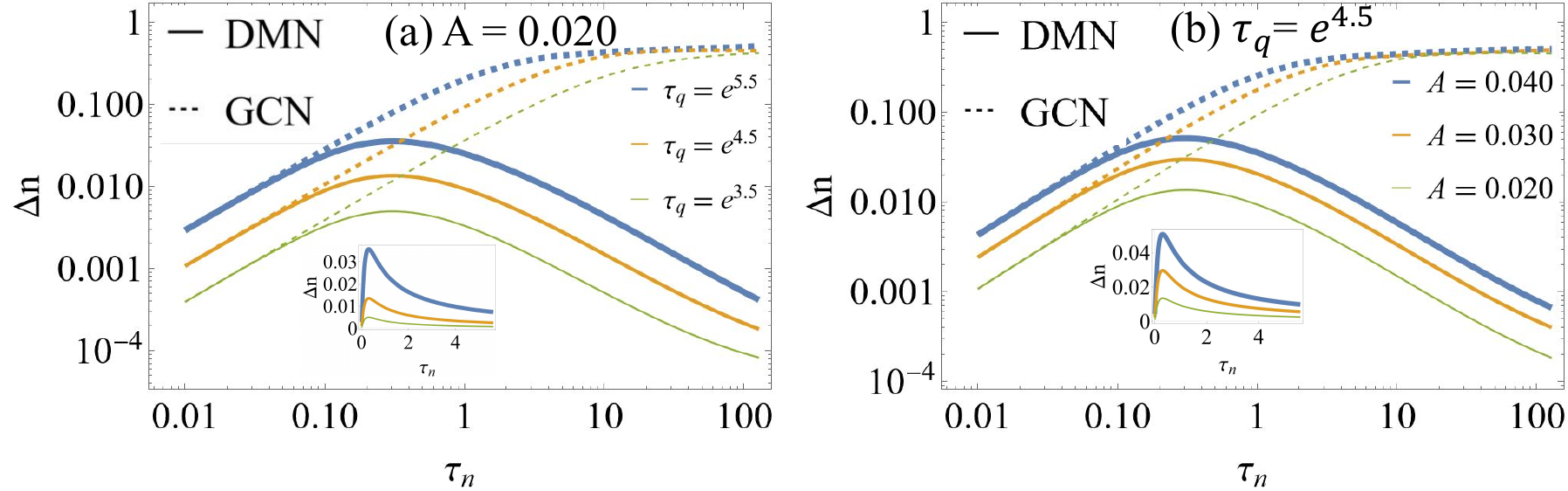}
		\caption{The dependence of the noise-induced defect density \(\Delta n\) on the correlation time \(\tau_n\) is compared for DMN (solid lines) and GCN (dashed lines). DMN depends nonmonotonically on \(\tau_n\): at fixed total noise power $A^2$ and quench time \(\tau_q\), it agrees with GCN only in the fast-noise regime, while beyond this regime \(\Delta n\) decreases as \(\tau_n\) increases. By contrast, \(\Delta n\) for GCN increases monotonically with \(\tau_n\) and saturates in the slow-noise regime, revealing distinct trends between the two noises.}\label{fig:3}
	\end{figure*}
Although DMN shares a Lorentzian power spectrum with GCN, the AKZ behaviors they induce differ considerably. As shown in Fig.~\ref{fig:3}, DMN yields a nonmonotonic noise-induced defect density, whereas GCN yields a monotonic one. To understand their difference, we analyze and compare the AKZ behaviors driven by DMN and GCN. In the following, we first review the AKZ behavior driven by GCN, and then focus on that driven by DMN.

\subsection{AKZ behavior driven by GCN}
\begin{figure*}[htbp]
		\centering
		\includegraphics[width=0.98\linewidth]{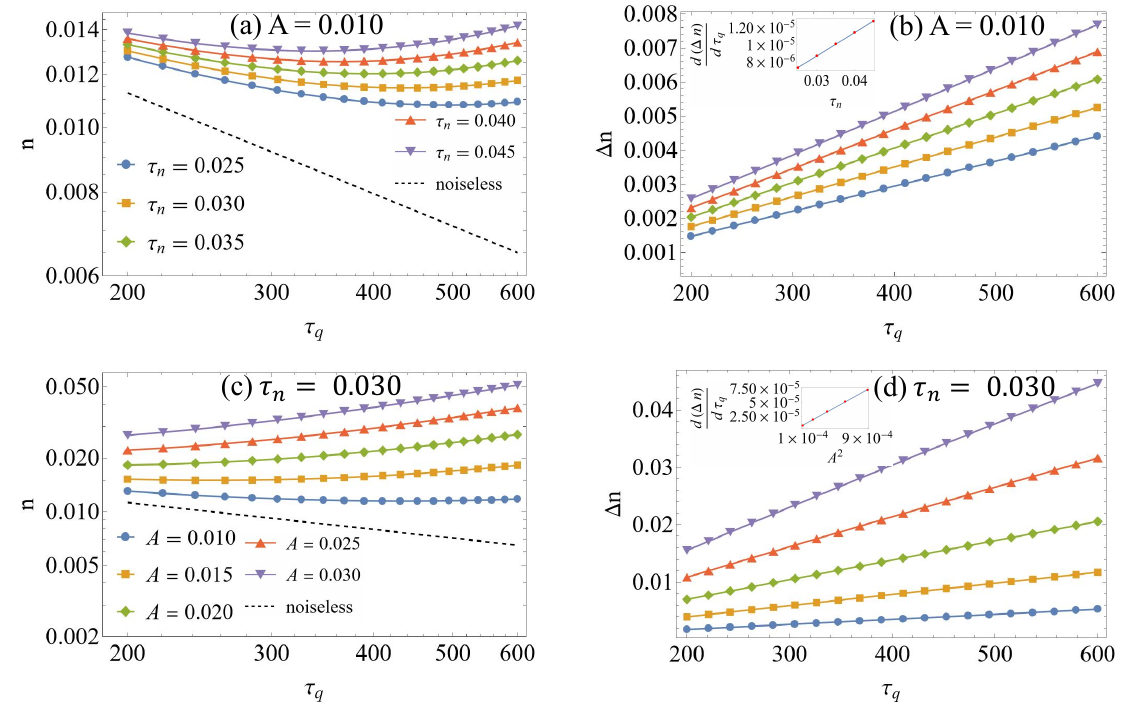}
		\caption{Figures~\ref{fig:1}(a) and~\ref{fig:1}(c) show how the AKZ behavior driven by DMN in the fast-noise regime varies with the noise correlation time $\tau_n$ and the noise intensity $A^2$, respectively. Likewise, Figs.~\ref{fig:1}(b) and~\ref{fig:1}(d) show the linear dependence of the slope $d\Delta n/d\tau_q$ on the corresponding noise parameter.}\label{fig:1}
	\end{figure*}

	Following Ref.~\cite{Sadeghizade2025}, the defect density induced by GCN is assumed to be proportional to the average noise power $S_{\mathrm{av}}$ over the frequency interval $|\omega|<\omega_c$, i.e., \begin{equation}S_{\mathrm{av}} = \frac{1}{2\omega_c} \int_{-\omega_c}^{\omega_c} S(\omega)\,d\omega,	\label{eq:average_noise_power}\end{equation} where $\omega_c=c/\tau_q$ is the characteristic quench frequency and $c$ is a  constant that can only be determined numerically. Consequently, through the power spectrum $S(\omega)$, the correlation time $\tau_n$ can strongly affect the noise-induced defect density $\Delta n$. This is demonstrated by GCN with an Ornstein--Uhlenbeck correlation, for which the Lorentzian power spectrum~\eqref{spec_gauss} yields \begin{equation}\Delta n \propto S_{\mathrm{av}} \tau_q = A^2 \tau_q^2 \arctan(c\tau_n/\tau_q).	\label{eq:defect_density}\end{equation} Clearly, $\Delta n$ increases monotonically with $\tau_n$.
This monotonic increase is characterized by two asymptotic regimes. In the fast-noise regime ($\tau_n\ll\tau_q$), $\Delta n \sim A^2 \tau_q \tau_n$ captures only a narrow low-frequency contribution while DMN follows this scaling consistently (Fig.~\ref{fig:1}). In the slow-noise regime ($\tau_n\gg\tau_q$), the scaling becomes $\Delta n \sim A^2 \tau_q^2$ where the entire noise power falls below $\omega_c$~\cite{Sadeghizade2025}.
	
	A direct inspection of Eq.~\eqref{eq:defect_density} reveals that the defect density induced by GCN is essentially dictated by the power spectrum. Because the statistical properties of Gaussian noise are entirely set by the power spectrum, the  AKZ behavior is also predominantly shaped by the power spectrum.
	\subsection{Generalized stochastic resonance in the AKZ behavior driven by DMN}
   \begin{figure*}[htbp]
		\centering
		\includegraphics[width=0.98\linewidth]{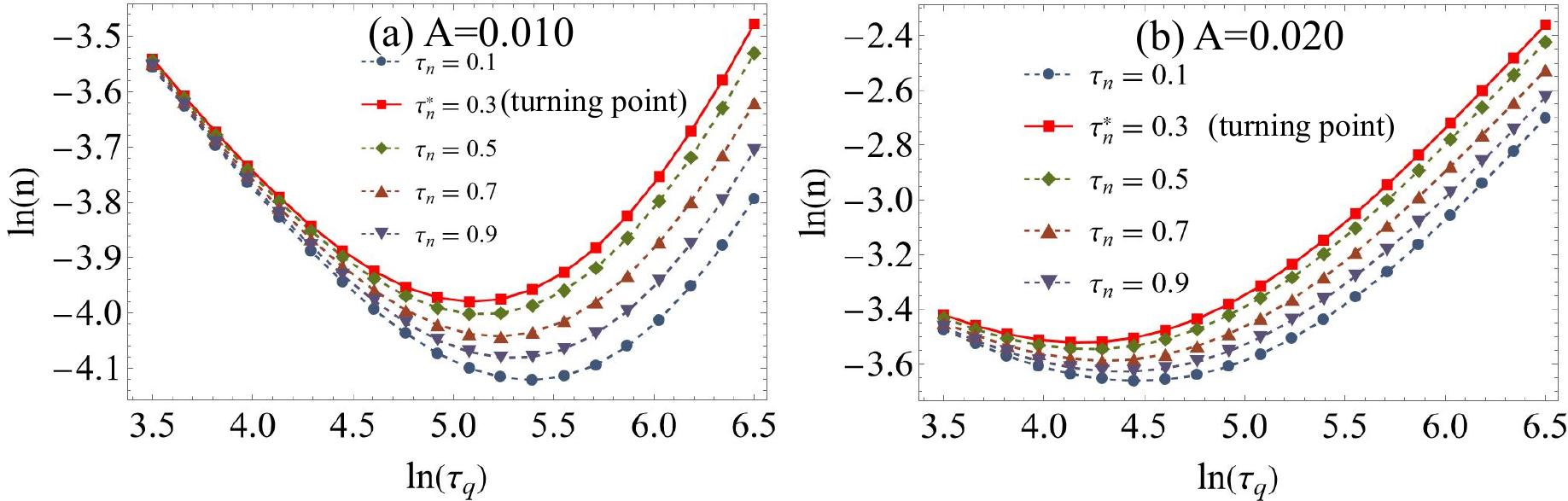}
		\caption{The DMN-driven AKZ behavior is enhanced and then suppressed as the correlation time $\tau_n$ grows. For weak noise power $A^2$, the enhancement persists below the turning point $\tau_n^*\approx 0.3$ (red line), whereas suppression occurs above $\tau_n^*$.}\label{fig:5}
	\end{figure*}	
	\begin{figure*}[htbp]
		\centering
		\includegraphics[width=0.98\linewidth]{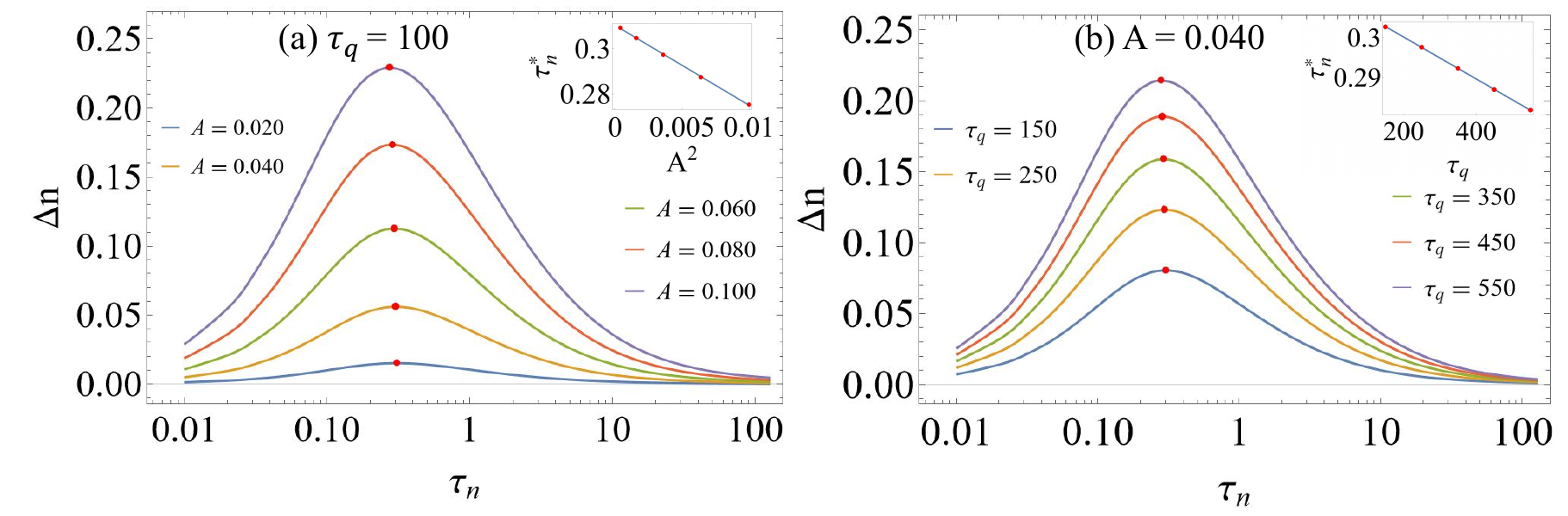}
		\caption{Figure~\ref{fig:4}(a) shows the dependence of the turning-point correlation time $\tau_n^*$ on the noise intensity $A^2$ at fixed quench time $\tau_q$, with red dots marking $\tau_n^*$. Figure~\ref{fig:4}(b) presents the corresponding dependence of $\tau_n^*$ on the quench time $\tau_q$ at fixed $A^2$.}\label{fig:4}
	\end{figure*}
	\begin{figure*}[htbp]
		\centering
		\includegraphics[width=0.98\linewidth]{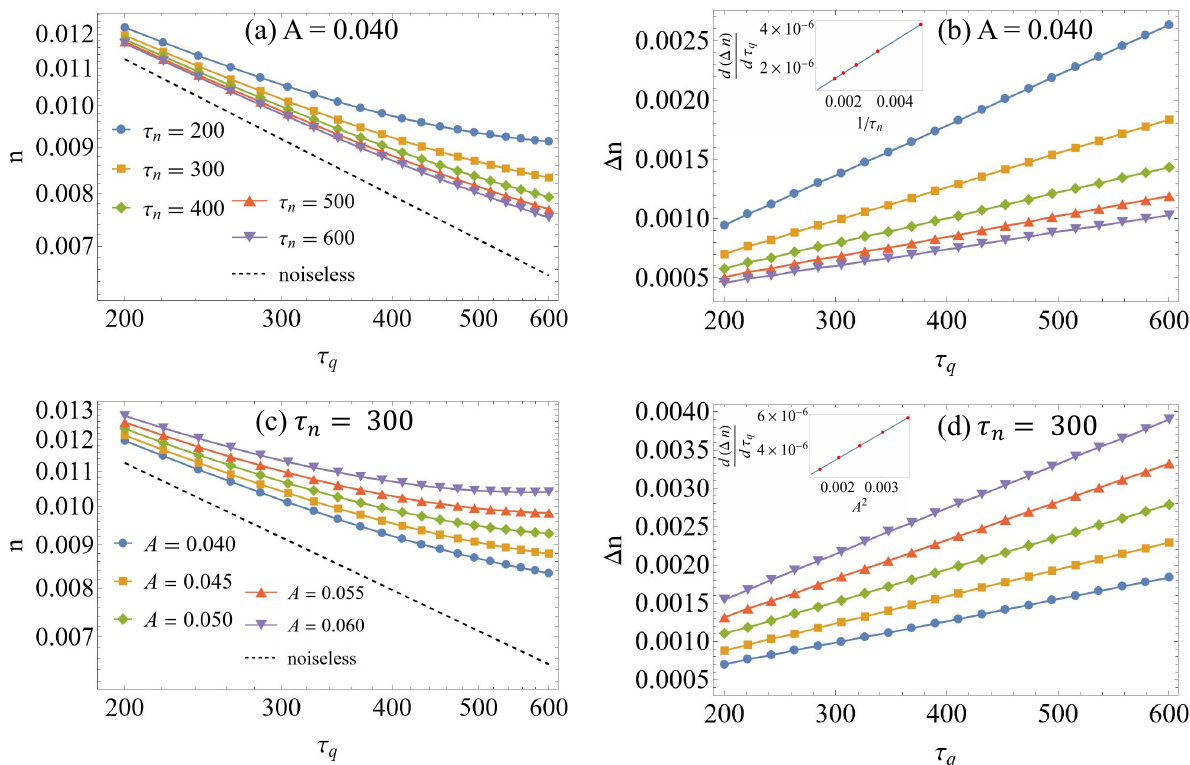}
		\caption{Figures~\ref{fig:2}(a) and~\ref{fig:2}(c) show how the AKZ behavior driven by DMN in the slow-noise regime varies with the noise correlation time $\tau_n$ and the noise intensity $A^2$, respectively. Correspondingly, Figs.~\ref{fig:2}(b) and~\ref{fig:2}(d) show the linear dependence of the slope $d\Delta n/d\tau_q$ on $1/\tau_n$ and $A^2$.}\label{fig:2}
	\end{figure*}
Unlike GCN, which yields a monotonic noise-induced defect density with respect to the correlation time \(\tau_n\), DMN yields a nonmonotonic one. Consequently, \(\Delta n\) reaches a maximum at a particular correlation time, which we term the turning-point correlation time \(\tau_n^*\). The AKZ behavior is therefore enhanced below the \(\tau_n^*\) and diminished above it (Fig.~\ref{fig:5}). Figure~\ref{fig:4}(a) shows that \(\tau_n^*\) decreases linearly with increasing  noise power \(A^2\) in the weak-noise regime. Figure~\ref{fig:4}(b) further shows that \(\tau_n^*\) decreases slowly with increasing quench time \(\tau_q\), with a linear dependence in the slow-quench regime.

The nonmonotonic dependence of $\Delta n$ on $\tau_n$ can be regarded as a GSR in the DMN-driven AKZ behavior, with $\tau_n$ as the noise parameter and $\Delta n$ as the response function. This aligns with the defining characteristic of GSR, which manifests as non-monotonic response behavior under varying noise parameters~\cite{Tian2025,Tian2026}; here, $\tau_n$ plays exactly this role. Furthermore, since GSR does not necessitate a periodic deterministic input~\cite{Collins1996}, the linear quench protocol in Eq.~\eqref{17} serves as the non-periodic deterministic drive in our framework.

\subsection{Sparse realization of DMN in the slow-noise regime}
The above GSR is a consequence of the DMN-driven AKZ behavior decreasing with $\tau_n$ in the slow-noise regime, a trend exactly opposite to that of GCN and thus anomalous. As further shown in Fig.~\ref{fig:2}, this decrease follows the scaling $\Delta n \sim A^2 \tau_q / \tau_n$.

  To analyze this anomalous AKZ behavior, it is necessary to begin with the detailed statistical properties of DMN.  According to Eqs.~\eqref{eq:POisson} and \eqref{correlation_time}, DMN is generated by a two-sided Poisson process \(N(t)\) of rate \(1/(2\tau_n)\).
Using Eq.~\eqref{eq:B7} in the slow-noise regime, we approximate the probabilities of zero and one switches of the DMN within the Landau--Zener interval $[-\tau_{\mathrm{LZ}}^{(k)}, \tau_{\mathrm{LZ}}^{(k)}]$ where the transitions mainly occur as
	\begin{gather}
		P(N(2\tau_{\mathrm{LZ}}^{(k)})=0)\simeq 1-\tau_{\mathrm{LZ}}^{(k)}/\tau_n,\\
		P(N(2\tau_{\mathrm{LZ}}^{(k)})=1)\simeq \tau_{\mathrm{LZ}}^{(k)}/\tau_n.
		\label{eq:1}
	\end{gather}
respectively.
	
	This result explicitly reveals the quasi-static nature of the noise. Specifically, $P(N(2\tau_{\mathrm{LZ}}^{(k)})=0)$ is close to unity, indicating that the DMN remains in its initial state almost surely over the Landau--Zener interval $[-\tau_{\mathrm{LZ}}^{(k)}, \tau_{\mathrm{LZ}}^{(k)}]$. The small probability \(P(N(2\tau_{\mathrm{LZ}}^{(k)})=1)\) corresponds to the {\it sparse realizations} involving a single random switch, while events with two or more switches are of order \((\tau_{\mathrm{LZ}}^{(k)} / \tau_n)^2\) and are therefore subleading.
Hence, the {\it sparse realizations} of DMN dominate the noise-induced transition probability $\Delta n^{(k)}$. Since the switching probability in Eq.~\eqref{eq:1} scales as $\tau_{\mathrm{LZ}}^{(k)}/\tau_n$, it follows that
\begin{equation}
\Delta n^{(k)} \sim \tau_{\mathrm{LZ}}^{(k)} / \tau_n,
\label{eq:NITP}
\end{equation}
which is consistent with~\cite{Luo2017,Sandeep2026}.
	\begin{figure}[htbp]
		\centering
		\includegraphics[width=0.98\linewidth]{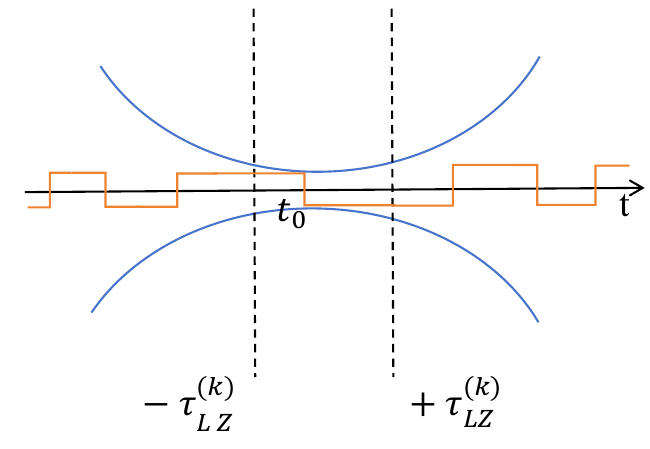}
		\caption{Schematic illustration of the {\it sparse realizations} of DMN. The blue curves denote the Landau--Zener adiabatic energy levels, and the yellow random square wave represents the DMN. A {\it sparse realization} corresponds to a single random switch occurring at $t_0$ within the Landau--Zener interval $[-\tau_{\mathrm{LZ}}^{(k)}, \tau_{\mathrm{LZ}}^{(k)}]$.}\label{fig:6}
	\end{figure}	
	
    As follows from Eqs.~\eqref{eq:HAM} and \eqref{eq:REP2}, the dynamics generated by the stochastic Hamiltonian~\eqref{ISING} reduces to a collection of decoupled noisy Landau--Zener processes corresponding to different $k$ modes.  The noise-induced defect density $\Delta n$ is therefore the average over modes of the noise-induced transition probabilities $\Delta n^{(k)}$ of these processes. Combining this with Eqs.~\eqref{eq:LZ} and \eqref{eq:NITP}, we find that \(\Delta n\) scales as \(\tau_q/\tau_n\) (Fig.~\ref{fig:2}(b)), consistent with the analytical derivation in Ref.~\cite{Sandeep2026}. As a result, it is the sparse realizations of DMN that give rise to the anomalous AKZ behavior.
	
		\subsection{Non-Gaussian statistics beyond the power spectrum}
	In the slow-noise regime, the {\it sparse realizations} of DMN yield anomalous AKZ behavior. This behavior is not accessible from the power spectrum alone, since it originates from the Poissonian arrival statistics of the switching events inherent to DMN. Such an insufficiency of the power spectrum is echoed in the result of Ref.~\cite{Rossi2016} that the different statistical properties of DMN and GCN directly influence the dynamics of the qubit system even when the two noises have identical power spectra. It also parallels the classic finding that the spectral response of a quadratic device to non-Gaussian noise is not determined by the power spectrum alone~\cite{Magness1954}.
	
	Thus an analysis based solely on the power spectrum may not always be adequate for systems subject to non-Gaussian noise such as DMN. Instead, attention may be better directed to the statistical details of the noise. The anomalous AKZ behavior illustrates this point: different noise types lead to contrasting physical behaviors particularly for Gaussian and non-Gaussian noise.
	 
    \section{summary}\label{sum}
    Focusing on the AKZ behavior, we investigate the role of DMN, a prototypical non-Gaussian noise, in the quantum critical dynamics of the one-dimensional transverse-field Ising chain. In the fast-noise regime, we find $\Delta n \sim A^2 \tau_n \tau_q$, a scaling similar to that of Gaussian noise. In the slow-noise regime, $\Delta n \sim A^2 \tau_q/\tau_n$, a behavior originating from the {\it sparse realizations} of DMN that eludes the description of the power spectrum. In stark contrast to GCN, the correlation time $\tau_n$ for DMN plays opposite roles in the two regimes. This opposite role gives rise to the nonmonotonic dependence of $\Delta n$ on $\tau_n$, manifesting as a generalized stochastic resonance (GSR). Our results demonstrate that for non-Gaussian noise, detailed statistical properties are essential for characterizing the AKZ behavior, thereby establishing a fundamental distinction between Gaussian and non-Gaussian noises in quantum critical dynamics.
\section*{ACKNOWLEDGMENTS}
We thank Dibyajyoti Sahu, Guglielmo Mazzola, Alireza Akbari, and Yan He for their interest in and discussions about this work.
\section*{DATA AVAILABILITY}
The data and software that support this manuscript are not publicly available. Requests for further information or access to the data should be directed to the authors.	
	\appendix
	
	\section{Independent mode dynamics}
	\label{app:derivation}
	
	In the tensor-product space $\mathcal{H} = \bigotimes_{k>0} \mathbb{C}^2$, 
	the Hamiltonian takes the form $H_\eta(t) = \sum_{k>0} H_{k,\eta}(t)$, 
	where each $H_{k,\eta}(t) = \Psi_{k}^{\dagger} h_{k,\eta}(t) \Psi_{k}$ 
	is a local quadratic operator acting on mode $k$~\eqref{DT}.  
	Consequently, different modes commute at all times:
	\begin{equation}
		[H_{k,\eta}(t), H_{k',\eta}(s)] = 0 \qquad (\forall\, k \neq k',\ \forall\, t,s). \label{eq:commute}
	\end{equation}
	
	By Eq.~\eqref{eq:commute}, the time-ordered exponential factorizes into a tensor product over the modes:
	\begin{equation}
		\begin{split}
			U_\eta(t) &= \mathcal{T}\exp\!\left[-i\int_0^t H_\eta(s)ds\right]
			= \bigotimes_{k>0} U_{k,\eta}(t) \\
			&= \bigotimes_{k>0} \mathcal{T}\exp\!\left[-i\int_0^t H_{k,\eta}(s)ds\right].
		\end{split}
	\end{equation}
	Given the initial product state $\rho_\eta(0) = \bigotimes_{k>0} \rho_k(0)$,
	the time-evolved density matrix retains the product form:
	\begin{equation}
		\rho_\eta(t) = \bigotimes_{k>0} \bigl[ U_{k,\eta}(t) \rho_k(0) U_{k,\eta}^\dagger(t) \bigr]
		= \bigotimes_{k>0} \rho_{k,\eta}(t).
		\label{app:rho_prod}
	\end{equation}
	
	The partial trace $\Tr_{\neq k}$ over all modes except $k$ is defined for any product
	operator $A\bigotimes B$ acting on $\mathcal{H}$,
	where $A$ lives on $\mathcal{H}_k$ and $B$ lives on the orthogonal
	complement $\mathcal{H}_k^\perp = \bigotimes_{j\neq k}\mathcal{H}_j$, as
	$\Tr_{\neq k}(A\bigotimes B) = A \, \Tr(B)$.   
	Applying the partial trace $\Tr_{\neq k}$ to Eq.~\eqref{app:rho_prod} gives
	\begin{equation}
		\Tr_{\neq k}\bigl( \rho_{\eta}(t) \bigr) = \rho_{k,\eta}(t).
		\label{app:pt_action}
	\end{equation}
	Using this result and the product form~\eqref{app:rho_prod}
	in Eq.~\eqref{eq:FN}, taking $\Tr_{\neq k}$ yields
	\begin{equation}
		{\frac{d}{dt}\rho_{k,\eta}(t) = -i [H_{k,\eta}(t), \rho_{k,\eta}(t)]}.
		\label{app:final}
	\end{equation}

	\section{Statistical properties of DMN }
	\label{app:FCS}
	
	Let $\eta(t)\in\{+A,-A\}$ be a symmetric dichotomous Markov noise, i.e., a continuous-time Markov process with switching rate $\lambda$:
	\begin{equation}
		\mathrm{P}\bigl(\eta(t+dt)=-A \mid \eta(t)=+A\bigr) = \lambda\,dt.
		\label{eq:B1}
	\end{equation}
	For $t>0$, define counting process  $N(t)$ as the number of sign changes in the time interval $[0,t]$, and let $P_{a,n}(t)$ be the joint probability that $\eta(t)=a$ and $N(t)=n$, where $a\in\{+A,-A\}$.
	The master equations yield:
	\begin{subequations}\label{eq:B2}
		\begin{align}
			\frac{d}{dt} P_{+,n} &= -\lambda P_{+,n} + \lambda P_{-,n-1},\\
			\frac{d}{dt} P_{-,n} &= -\lambda P_{-,n} + \lambda P_{+,n-1}.
		\end{align}
	\end{subequations}
	with $P_{\pm,-1}\equiv0$. 
	Introduce the conditional characteristic functions $G_a(\mu,t)=\sum_{n=0}^\infty P_{a,n}(t)e^{i\mu n}$, 
	which obey
	\begin{subequations}\label{eq:B3}
		\begin{align}
			\frac{d}{dt} G_+ &= -\lambda G_+ + \lambda e^{i\mu}G_-,\\
			\frac{d}{dt} G_- &= -\lambda G_- + \lambda e^{i\mu}G_+.
		\end{align}
	\end{subequations}
	Solving \eqref{eq:B3} with the stationary initial condition $P_\pm(0)=1/2$ yields the 
	unconditional characteristic function $G(\mu,t) = \frac{1}{2}G_{+} + \frac{1}{2}G_{-}$ as
	\begin{equation}
		G(\mu,t) = e^{-\lambda  t(1-e^{i\mu})}.
		\label{eq:B5}
	\end{equation}
	Therefore, by symmetry, the case $t<0$ follows analogously, and by the uniqueness of the characteristic function, $N(t)$ follows a Poisson distribution with parameter $\lambda |t|$:
	\begin{equation}
		P(N(t)=N) = e^{-\lambda |t|}\frac{(\lambda |t|)^N}{N!}.
		\label{eq:B7}
	\end{equation}
	The counting process $N(t)$ is extended to all $t$ by taking the interval $[t,0)$ for $t<0$ and $(0,t]$ for $t>0$; it therefore depends only on $|t|$. Consequently, from Eq.~\eqref{eq:B7}, $N(t)$ is a two-sided Poisson process of rate $\lambda$, which yields the full counting statistics of $N(t)$.
	Hence, the symmetric dichotomous Markov noise $\eta(t)$ becomes
	\begin{equation}
		\eta(t) = \eta(0)(-1)^{N(t)},
		\label{eq:poisson}
	\end{equation}
	with the stationary initial condition \begin{equation}P(\eta(0) = +A) = P(\eta(0) = -A) = 1/2.\label{eq:stationary initial condition}
	\end{equation}
	
	Because the noise $\eta(t)$ is symmetric, it is valid to take $\eta(0)=+A$, as guaranteed by Eq.~\eqref{eq:stationary initial condition}. For simplicity, consider $t>0$.
	It then follows that $\eta(t)=+A$ if and only if the switching count is even, 
	and $\eta(t)=-A$ if and only if it is odd.
	Thus, from this correspondence and the Eq.~\eqref{eq:B7}, the 
	transition probabilities follow as
	\begin{equation}
		\begin{aligned}
			P_{++}(t) &\equiv \mathrm{P}\bigl(\eta(t)=+A \mid \eta(0)=+A\bigr) \\
			&= e^{-\lambda t}\sum_{k=0}^{\infty}\frac{(\lambda t)^{2k}}{(2k)!}
			= \frac{1+e^{-2\lambda t}}{2},
		\end{aligned}
		\label{eq:B8}
	\end{equation}
	
	\begin{equation}
		\begin{aligned}
			P_{+-}(t) &\equiv \mathrm{P}\bigl(\eta(t)=-A \mid \eta(0)=+A\bigr) \\
			&= e^{-\lambda t}\sum_{k=0}^{\infty}\frac{(\lambda t)^{2k+1}}{(2k+1)!}
			= \frac{1-e^{-2\lambda t}}{2}.
		\end{aligned}
		\label{eq:B9}
	\end{equation}
	Therefore, the transition matrix is
	\begin{equation}
		\begin{aligned}
			T(t) &\equiv \begin{pmatrix}P_{++}(t)&P_{+-}(t)\\P_{-+}(t)&P_{--}(t)\end{pmatrix} \\
			&= \frac{1}{2}\begin{pmatrix}1+e^{-2\lambda t}&1-e^{-2\lambda t}\\1-e^{-2\lambda   t}&1+e^{-2\lambda t}\end{pmatrix},
		\end{aligned}
		\label{eq:B10}
	\end{equation}
	which satisfies the Chapman--Kolmogorov equation:
	\begin{equation}
		T(t+s) = T(t)T(s), \qquad T(0)=I.		\label{eq:B11}
	\end{equation}
	In turn, the correlation function follows directly from the transition 
	probabilities, which are specified in Eqs.~\eqref{eq:B8} and \eqref{eq:B9}:
	\begin{equation}
		\langle \eta(0)\eta(t) \rangle = A^2\bigl[P_{++}(t) - P_{+-}(t)\bigr] = A^2e^{-2\lambda t},\; \forall t>0,
		\label{eq:B12}
	\end{equation}
	which yields the correlation time
	\begin{equation}
		\tau_n = \int_0^\infty \frac{\langle \eta(t)\eta(0) \rangle}{\langle \eta(0)^2 \rangle}\,dt = \int_0^\infty e^{-2\lambda t}\,dt = \frac{1}{2\lambda}.
		\label{eq:B13}
	\end{equation}
	With the symmetry for $t<0$ included, inserting the correlation time defined in Eq.~\eqref{eq:B13} into Eq.~\eqref{eq:B12} yields
	\begin{equation}
		R(t) = \langle \eta(0)\,\eta(t) \rangle = A^{2} \exp\!\left(-\frac{|t|}{\tau_{n}}\right).
		\label{eq:corr}
	\end{equation}

\bibliography{references}	
\end{document}